\documentstyle[pre,aps,amsfonts,amssymb,twocolumn,eqsecnum]{revtex}
\begin{document}
\def\phi{\varphi} \def\epsilon{\varepsilon}
\def\u{\bbox}

\draft \title{Approximate renormalization for the breakup of invariant tori
  with three frequencies}

\author{C.\ Chandre$^1$ and R.S. MacKay$^2$} 
\address{$^1$Service de Physique Th\'eorique, CEA Saclay, F-91191
  Gif-sur-Yvette Cedex, France} 
\address{$^2$Mathematics Institute,
  University of Warwick, Coventry CV4 7AL, U.K.}
\maketitle

\begin{abstract}
  We construct an approximate renormalization transformation for
  Hamiltonian systems with three degrees of freedom in order to study
  the break-up of invariant tori with three incommensurate frequencies
  which belong to the cubic field $Q(\tau)$, where
  $\tau^3+\tau^2-2\tau-1=0$. This renormalization has two fixed
  points~: a stable one and a hyperbolic one with a codimension one
  stable manifold. We compute the associated critical exponents that
  characterize the universality class for the break-up of the
  invariant tori we consider.
\end{abstract}

\section{Introduction}\label{sec.1}

In this paper, we define an approximate renormalization scheme for
Hamiltonians with three degrees of freedom in order to study the
break-up of invariant tori with the frequency vector
$\u{\omega}_0=(\tau^2+\tau,\tau,1)$ where $\tau=2\cos(2\pi/7)\approx
1.2469796$ is the root of modulus larger than one of the polynomial~:
$$
\tau^3+\tau^2-2\tau-1=0.
$$
This approximate renormalization is defined for the following
family of Hamiltonians which are quadratic in the actions
$\u{A}=(A_1,A_2,A_3)$~:
\begin{equation}
  \label{eq:Hp}
  H(\u{A},\u{\phi})=H_0(\u{A})+V(\u{A},\u{\phi}),
\end{equation}
where the Hamiltonian $H_0$ is quadratic~:
\begin{equation}
  \label{eq:H0}
  H_0(\u{A})=\u{\omega}_0\cdot\u{A}+\frac{1}{2}(\u{\Omega}\cdot \u{A})^2,
\end{equation}
and the perturbation is described by two scalar functions of the
angles $\u{\phi}=(\phi_1,\phi_2,\phi_3)$~:
\begin{equation}
  \label{eq:V}
  V(\u{A},\u{\phi})=g(\u{\phi})\u{\Omega}\cdot\u{A}+f(\u{\phi}).
\end{equation}
For $H_0$, the invariant torus with frequency vector $\u{\omega}_0$ is
located at $\u{A}$ such that $\u{\Omega}\cdot\u{A}=0$ and
$\u{\omega}_0\cdot\u{A}=E$ where $E$ is the total energy of the
system. Since $\u{\omega}_0$ satisfies a diophantine condition, the
KAM theorem for Hamiltonians~(\ref{eq:Hp}) states that for a
sufficiently small and smooth perturbation $V$, this invariant torus
persists~\cite{chan98c}. For a sufficiently large perturbation, this
invariant torus is broken (converse KAM~\cite{mack85}).\\
The purpose of this paper is to construct a renormalization
transformation in order to investigate the properties of invariant
tori with the frequency vector $\u{\omega}_0$ at criticality.  The
idea of the renormalization approach is to iterate a transformation in
the space of Hamiltonians. For Hamiltonians that have a smooth
invariant torus with frequency vector $\u{\omega}_0$, the iteration
should converge to a trivial fixed point. The set of Hamiltonians that
have a non-smooth invariant torus with this frequency vector form a
surface (called critical surface) which is invariant under the
renormalization, and is expected to be a codimension one stable
manifold
of a non-trivial fixed set of the renormalization. \\
The aim is to define a renormalization as  
was done for
Hamiltonians with two degrees of freedom in
Refs.~\cite{koch99,chan98b,abad98}. This renormalization is a
combination of a partial elimination of the perturbation (we eliminate
the Fourier modes of the perturbation which are sufficiently far from
resonance) and a rescaling transformation which 
consists of a
shift of the resonances, and a rescaling of time and of the actions. \\
An approximate renormalization scheme for the frequency vector
$\u{\omega}_0$ was defined in Ref.~\cite{meis99} following the
renormalization defined for the spiral mean torus constructed in
Ref.~\cite{mack94}. The main result of Ref.~\cite{meis99} was that the
renormalization has a fixed point on the critical surface but it is
not hyperbolic.
This renormalization dynamics is structurally unstable.\\
We propose to 
improve on Ref.~\cite{meis99}, just as was done for Ref.~\cite{mack94} 
in Refs.~\cite{chan98d,chan99b} for the spiral mean torus. The
main difference with the scheme constructed in Ref.~\cite{meis99} is
that we use a different normalization condition for the rescaling in
the actions
and we include a term that was previously neglected. 
The main result we find is that a large part of the critical surface is
the codimension-one stable manifold of a hyperbolic fixed point of this
renormalization. We expect the dynamics of the exact renormalization
to be the same as this approximate renormalization, at least locally.\\
Following the approach of Escande and Doveil~\cite{esca81,esca85} for
systems with two
degrees of freedom, we perform two approximations in this renormalization~:\\
$(1)$ A three resonance approximation~: we keep only the three main
Fourier modes of the perturbation $V$ at each step of the
transformation.\\
$(2)$ A mean-value quadratic approximation~: we neglect the dependence
on the angles of the quadratic term in the actions.\\

The frequency vector $\u{\omega}_0$ is an eigenvector of the matrix
$\tilde{N}$ with the eigenvalue $(\tau +1)^{-1}\approx 0.445$, where
$\tilde{N}$ is the transposed matrix of the following matrix $N$ with
integer coefficients and determinant -1~:
\begin{equation}
  \label{eq:N}
  N=\left( \begin{array}{ccc} 
      0 & 1 & -1\\
      1 & -1 & 1\\
      0 & -1 & 2 
      \end{array}\right).
\end{equation}
From $\tilde{N}$ one generates a sequence of periodic orbits (with frequency
vector $\{ \u{\Omega}_k\}$) approximating the motion with frequency
vector $\u{\omega}_0$, i.e.\ $\u{\phi}(t)=\u{\omega}_0 t +\u{\phi}_0
\mbox{ mod } 2\pi$~:
$$
\u{\Omega}_k=(p_k/r_k,q_k/r_k,1)\rightarrow_{n\to \infty}
\u{\omega}_0,
$$
where $p_k$, $q_k$ and $r_k$ are determined by the following
recursion relations~:
\begin{eqnarray*}
  && p_{k+1}=p_k+2q_k+r_k,\\
  && q_{k+1}=p_k,\\
  && r_{k+1}=q_k+r_k,
\end{eqnarray*}
and $p_0=3$, $q_0=1$, $r_0=1$, i.e.\ 
$\u{\Omega}_k=\tilde{N}^{-k}\u{\Omega}_0$.  These relations define a
sequence of simultaneous rational approximations $(p_k/r_k,q_k/r_k)$
to the pair
$(\tau^2+\tau,\tau)$.\\
The sequence of resonances is generated by the matrix $N$ from the
vector $\u{\nu}_1=(1,0,0)$~:
$$
\u{\nu}_k=N^{k-1}\u{\nu}_1.
$$
The small denominators
decrease geometrically to zero with the ratio $(\tau +1)^{-1}\approx
0.445$~:
$$
\u{\omega}_0\cdot\u{\nu}_k=\tau(\tau +1)^{2-k}.
$$
The modes $\u{\nu}_k$ are in resonance with the periodic motion
with frequency vector $\u{\Omega}_{k}$ (with period $r_k$)~:
$$
\u{\Omega}_k\cdot \u{\nu}_{k+4}=0.
$$
The matrix $N$ can be decomposed in the following way~:
\begin{equation}
  \label{eq:LPLL}
  N=L P L^2,
\end{equation}
where
\begin{equation}
  \label{eq:L}
  L=\left( \begin{array}{ccc} 
      0 & 0 & 1\\
      1 & 0 & 0\\
      0 & 1 & -1 \end{array} \right),
\end{equation}
and
\begin{equation}
  \label{eq:P}
  P=\left( \begin{array}{ccc} 
      0 & 0 & 1\\
      0 & 1 & 0\\
      1 & 0 & 0 \end{array} \right).
\end{equation}
We notice that $\mbox{det }L=-\mbox{det }P=1$ and that $P$ is a
permutation of two elements of the basis; in particular, it satisfies
$P^2=1$. This decomposition follows from the construction of the Farey
sequence for an incommensurate vector $\u{\omega}\in {\Bbb
  R}^3$~\cite{kim86,mack94,meis99}. With this decomposition of the
matrix $N$, we construct the renormalization from two operators (one
associated with $L$ and the other one to $P$) in a similar way as MacKay
did for Hamiltonian systems with two degrees of freedom in
Ref.~\cite{mack88} in order to investigate the break-up of invariant
tori with arbitrary winding ratio. The approximate renormalization
transformation we define for $\u{\omega}_0$ can be constructed for a
more general frequency vector from these two operators given the Farey
decomposition of the frequency vector. For instance, concerning the
spiral mean torus~\cite{chan98d,chan99b}, the renormalization is equal
to one of the operators (the one which is associated with $L$).

\section{Definition of the renormalization transformation}

The renormalization transformation $\mathcal{R}$ we define for the
frequency vector $\u{\omega}_0$ is composed of two
operators~: one associated with the matrix $P$ and the other one with
$L$. We denote ${\mathcal{R}}_P$ and ${\mathcal{R}}_L$ these operators.
The transformation $\mathcal{R}$ is defined for a {\em fixed}
frequency vector $\u{\omega}_0$ and is given by
$$
{\mathcal{R}}={\mathcal{R}}_L \circ {\mathcal{R}}_L \circ {\mathcal{R}}_P
\circ {\mathcal{R}}_L.
$$

\subsection{Definition of ${\mathcal{R}}_P$}
\label{sec:RP}

The renormalization operator ${\mathcal{R}}_P$ acts on the following
Hamiltonians~:
\begin{equation}
  \label{eq:hamrp}
  H(\u{A},\u{\phi})=H_0(\u{A})+h(\u{\Omega}\cdot \u{A},\u{\phi}),
\end{equation}
where
\begin{eqnarray*}
&& H_0(\u{A})=\u{\omega}\cdot\u{A}+\frac{1}{2}(\u{\Omega} \cdot 
  \u{A})^2,\\
&& h(\u{\Omega}\cdot\u{A},\u{\phi})=\sum_{i=1}^3 h_i(\u{A}) 
\cos\phi_i=\sum_{i=1}^3 (f_i+g_i\u{\Omega}
  \cdot \u{A})\cos\phi_i.
\end{eqnarray*}
It contains a shift of the Fourier modes and a rescaling of time and
of the actions. The shift of the Fourier modes is constructed such
that it exchanges the mode $(1,0,0)$ and $(0,0,1)$ without changing
the mode $(0,1,0)$. More precisely, we require that the new angles
$\u{\phi}'$ satisfy~:
\begin{eqnarray*}
  && \cos\phi'_1=\cos\phi_3,\\
  && \cos\phi'_2=\cos\phi_2,\\
  && \cos\phi'_3=\cos\phi_1.
\end{eqnarray*}
This is performed by the following linear canonical transformation~:
$$
(\u{A},\u{\phi})\mapsto (\u{A}',\u{\phi}')=(P\u{A},P\u{\phi}),
$$
where we recall that $P$ is symmetric and orthogonal. The vectors
$\u{\omega}$ and $\u{\Omega}$ are changed into $P\u{\omega}$ and
$P\u{\Omega}$. We impose that the images $\u{\omega}'$ and
$\u{\Omega}'$ of the vectors $\u{\omega}$ and $\u{\Omega}$ by the
renormalization ${\mathcal{R}}_P$ satisfy the following normalization
conditions~: $\u{\Omega}'$ must be of (euclidean) norm one, and the
third component of $\u{\omega}'$ must be equal to one. Since $P$ is
orthogonal, $P\u{\Omega}$ is again of norm one, and thus
$\u{\Omega}'=P\u{\Omega}$.  The third component of $P\u{\omega}$ is
equal to $\omega_1$. We rescale the time by a factor $\omega_1$, i.e.\ 
we multiply the Hamiltonian by $1/\omega_1$. Then $\u{\omega}'$ is
given by $\u{\omega}'=
\u{\omega}/\omega_1=(\omega_3/\omega_1,\omega_2/\omega_1,1)$.\\
The quadratic part of the Hamiltonian $H_0$ becomes
$(\u{\Omega}'\cdot\u{A})^2/(2\omega_1)$.  In order that this quadratic
term is equal to $(\u{\Omega}'\cdot\u{A})^2/2$, we rescale the actions
by a factor $\lambda_P=1/\omega_1$, i.e.\ we change the
Hamiltonian $H$ into $\lambda_P H(\u{A}/\lambda_P,\u{\phi})$.\\
In summary, a Hamiltonian $H$ given by Eq.~(\ref{eq:hamrp}) is mapped
into
$$
H'(\u{A},\u{\phi})=\u{\omega}'\cdot\u{A}+\frac{1}{2}(\u{\Omega}'
\cdot \u{A})^2+\sum_{i=1}^3
(f'_i+g'_i\u{\Omega}'\cdot\u{A})\cos\phi_i,
$$
where $\u{\omega}'=(\omega_3/\omega_1,\omega_2/\omega_1,1)$,
$\u{\Omega}'=(\Omega_3,\Omega_2,\Omega_1)$, and
\begin{eqnarray*}
  && f'_1=f_3/\omega_1^2,\\
  && f'_2=f_2/\omega_1^2,\\
  && f'_3=f_1/\omega_1^2,\\
  && g'_1=g_3/\omega_1,\\
  && g'_2=g_2/\omega_1,\\
  && g'_3=g_1/\omega_1.
\end{eqnarray*}
The renormalization operator ${\mathcal{R}}_P$ is equivalent to the
10-dimensional map given by the above equations (we recall that we
impose a normalization condition on $\u{\omega}$ and $\u{\Omega}$)~:
$$
(\u{\omega},\u{\Omega},f_i,g_i;i=1,2,3)\mapsto
(\u{\omega}',\u{\Omega}',f'_i,g'_i;i=1,2,3).
$$

\subsection{Definition of ${\mathcal{R}}_L$}
\label{sec:RL}

The renormalization operator ${\mathcal{R}}_L$ is associated to the
matrix $L$ and acts on the family of Hamiltonians~(\ref{eq:hamrp}). It
contains an elimination of the mode $\u{\nu}_1$ of the perturbation,
and a rescaling procedure (shift of the resonances, rescaling of time
and of the actions) such that the image of a Hamiltonian $H$ given by
Eq.~(\ref{eq:hamrp}) is of the same general form as $H$ and describes
the system on a smaller scale in phase space and at a longer time scale.
This operator is similar to the approximate renormalization
constructed for the spiral mean torus in
Refs.~\cite{chan98d,chan99b}.\\
We eliminate the mode $\u{\nu}_1$ of the scalar functions $f$ and $g$,
by a near-identity canonical transformation. We perform a Lie
transformation generated by a function $S(\u{A},\u{\phi})$. The image
of a Hamiltonian $H$ is given by~:
\begin{equation}
  H'=\exp(\hat{S})H=H+\{S,H\}+\frac{1}{2}\{S,\{S,H\}\}+\cdots,
\end{equation}
where $\{\, , \, \}$ denotes the Poisson bracket~:
$$
\{f,g\}=\frac{\partial f}{\partial \u{\phi}}\cdot \frac{\partial
  g}{\partial \u{A}} - \frac{\partial g}{\partial \u{\phi}}\cdot
\frac{\partial f}{\partial \u{A}},
$$
and the operator $\hat{S}$ acts on $H$ like $\hat{S}H=\{ S,H\}$.
The generating function is chosen linear in the actions and of the
form~:
\begin{equation}
  \label{eq:S1}
  S(\u{A},\u{\phi})=(z+y\u{\Omega}\cdot
  \u{A})\sin\phi_1.
\end{equation}
We choose $z$ and $y$ in the following way~:
\begin{eqnarray*}
  && z=-f_1/\omega_1,\\
  && y=-(g_1-f_1 \Omega_1/\omega_1)/\omega_1.
\end{eqnarray*}
Then the generating function $S$ satisfies~:
$$
\{ S,H_0\}=(Q(\u{A})-h_1(\u{A}))\cos\phi_1,
$$
with $Q(\u{A})=y\Omega_1 (\u{\Omega}\cdot\u{A})^2$. The image of a
Hamiltonian $H$ given by Eq.~(\ref{eq:hamrp}) is~:
\begin{eqnarray*}
  H'= && H_0+h_1 \cos\phi_1+h_2 \cos\phi_2 +h_3 \cos\phi_3\\
&& +\{S,H_0\}+\{S,h_1 \cos\phi_1 \}
   +\{S,h_2 \cos\phi_2\}\\
    &&  +\{S,h_3 \cos\phi_3\}+\frac{1}{2}\{S,\{S,H_0\}\}+O(\varepsilon^3).
\end{eqnarray*}
We neglect the term $Q(\u{A}) \cos\phi_1$ produced by the terms
$\{S,H_0\}+ h_1 \cos\phi_1$, i.e.\ we neglect the dependence on the
angles of the quadratic terms in the actions. The term $\{S,h_1
\cos\phi_1\}$ is of degree one in the actions and contains the modes
$\u{0}$ and $2\u{\nu}_1$. We neglect the mode $2\u{\nu}_1$ and we
eliminate the mode $\u{0}$ of the linear term in the actions by a
shift~: $\u{A}'=\u{A} +\u{a}$ where $\u{a}$ is of order
$\varepsilon^2$ (in the direction of $\u{\Omega}$).  We neglect the
modes with frequency vectors $\u{\nu}_1\pm \u{\nu}_2$ produced by the
term $\{S,h_2 \cos\phi_2\}$.  The term $\{S,h_3 \cos\phi_3\}$
generates the modes $\u{\nu}_1\pm \u{\nu}_3$. We neglect the mode
$\u{\nu}_1+\u{\nu}_3$ and keep the next relevant Fourier mode
$\u{\nu}_1-\u{\nu}_3$ whose amplitude is denoted
$h'_3=f'_3+g'_3\u{\Omega}\cdot \u{A}$, where
\begin{eqnarray}
  && f'_3=(z g_3\Omega_1+yf_3\Omega_3)/2,\label{eq:f'3}\\
  && g'_3=(\Omega_1+\Omega_3)y g_3/2. \label{eq:g'3}
\end{eqnarray}
The term $\{S,Q\cos\phi_1\}/2$ produced by $\{S,\{S,H_0\}\}/2$ gives a
contribution to the quadratic part of the Hamiltonian. The new
quadratic part is equal to $m(\u{\Omega}\cdot\u{A})^2/2$ where $m$ is
given by~:
\begin{equation}
  \label{eq:m}
  m=1+\frac{3}{2}y^2\Omega_1^2.
\end{equation}
Then after the elimination of the mode $\u{\nu}_1$, the new
Hamiltonian is equal to~:
\begin{eqnarray}
  H'=&&\u{\omega}\cdot \u{A}+ m(\u{\Omega}\cdot\u{A})^2/2+h_2\cos\phi_2
+h_3\cos\phi_3\nonumber \\
&& +h'_3\cos(\phi_1-\phi_3), \label{eq:h'}
\end{eqnarray}
where $h'_3$ is given by Eqs.~(\ref{eq:f'3}) and (\ref{eq:g'3}).\\
We shift the Fourier modes according to the following linear canonical
transformation~:
$$
(\u{A},\u{\phi})\mapsto
(\u{A}',\u{\phi}')=(L^{-1}\u{A},\tilde{L}\u{\phi}),
$$
such that $\cos\phi_2=\cos\phi'_1$, $\cos\phi_3=\cos\phi'_2$ and
$\cos(\phi_1-\phi_3)=\cos\phi'_3$. The vectors $\u{\Omega}$ and
$\u{\omega}$ are changed in the following way~:
\begin{eqnarray*}
  && \u{\Omega}'=\frac{\tilde{L}\u{\Omega}}{\Vert \tilde{L}\u{\Omega}
  \Vert}=(\Omega_2,\Omega_3,\Omega_1-\Omega_3)
           /(1+\Omega_3(\Omega_3-2\Omega_1)),\\
  && \u{\omega}'=\frac{\tilde{L}\u{\omega}}{(\tilde{L}\u{\omega})_3}
     =(\omega_2/(\omega_1-1),1/(\omega_1-1),1).
\end{eqnarray*}
We rescale time in such a way that the linear term in the actions of
$H_0$ is equal to $\u{\omega}'$~: we multiply the Hamiltonian $H'$ by
a factor $1/(\omega_1-1)$. The quadratic term of the new Hamiltonian
is equal to $m \Vert \tilde{L}\u{\Omega}\Vert^2
(\u{\Omega}'\cdot\u{A})^2/(2\omega_1-2)$. In order to map this
quadratic part into $(\u{\Omega}'\cdot\u{A})^2/2$, we rescale the
actions by a factor
$$
\lambda_L=m \Vert \tilde{L}\u{\Omega}\Vert ^2/(\omega_1-1),
$$
by changing the Hamiltonian $H'$ into $\lambda_L
H'(\u{A}/\lambda_L,\u{\phi})$.\\
After this elimination and rescaling procedures, a Hamiltonian $H$ is
mapped into
$$
H''=\u{\omega}'\cdot\u{A}+\frac{1}{2}(\u{\Omega}'\cdot\u{A})^2+
h''_1\cos\phi_1+ h''_2\cos\phi_2+h''_3\cos\phi_3,
$$
where $h''_i=f''_i+g''_i\u{\Omega}'\cdot\u{A}$ and
\begin{eqnarray*}
  && f''_1=f_2 c_f,\\
  && f''_2=f_3 c_f,\\
  && f''_3=f'_3 c_f,\\
  && g''_1=g_2 c_g,\\
  && g''_2=g_3 c_g,\\
  && g''_3=g'_3 c_g,
\end{eqnarray*}
with $c_f=m \Vert \tilde{L}\u{\Omega}\Vert ^2/(\omega_1-1)^2$
and $c_g= \Vert \tilde{L}\u{\Omega}\Vert /(\omega_1-1)$.\\
The renormalization transformation is equivalent to the 10-dimensional
map
$$
(\u{\omega},\u{\Omega},f_i,g_i;i=1,2,3)\mapsto
(\u{\omega}',\u{\Omega}',f''_i,g''_i;i=1,2,3),
$$
given by the above equations.

\subsection{Renormalization transformation $\mathcal{R}$}

In summary, the renormalization
${\mathcal{R}}={\mathcal{R}}_L^2{\mathcal{R}}_P {\mathcal{R}}_L$ acts in the
following way~: a canonical transformation eliminates the three main
Fourier modes and produce the next three resonances which are the main
Fourier modes at a smaller scale in phase space.  A rescaling
procedure shift these resonances into the original Fourier modes, and
normalize the new
Hamiltonian by rescaling the time and the actions.\\
The rescaling procedure acts on $\u{\omega}_0$ in the following way~:
$\u{\omega}_0$ is changed into
$\u{\omega}^{(1)}=(1/(3-\tau^2),1/(3\tau -\tau^3),1)$ by
${\mathcal{R}}_L$, then into $\u{\omega}_0^{(2)}=(3-\tau^2,\tau^{-1},1)$
by ${\mathcal{R}}_P$, then into
$\u{\omega}_0^{(3)}=(1/(2\tau-\tau^3),1/(2-\tau^2),1)$ by
${\mathcal{R}}_L$, and after the final step ${\mathcal{R}}_L$, into
$\u{\omega}_0^{(4)}=\u{\omega}_0$. Thus the renormalization
$\mathcal{R}$ is defined for a {\em fixed} frequency vector
$\u{\omega}_0$. In the same way, the action of $\mathcal{R}$ on
$\u{\Omega}$ reduces to the map~:
\begin{equation}
  \label{eq:Om}
  \u{\Omega}'=\frac{\tilde{N}\u{\Omega}}{\Vert
  \tilde{N}\u{\Omega}\Vert }.
\end{equation}
The spectrum of $\tilde{N}$ is real and 
consists of the following
eigenvalues $-\tau\approx -1.2469$, $(\tau+1)^{-1}\approx 0.4450$,
$1+\tau^{-1}\approx 1.8019$. Thus by iterating the map~(\ref{eq:Om}),
the vector $\u{\Omega}$ converges to the eigenvector (denoted
$\u{\Omega}^{(3)}$, with euclidean norm one) associated with the
largest eigenvalue of $\tilde{N}$. With this reduction, the
renormalization $\mathcal{R}$ reduces to a 6-dimensional map
$(f_i,g_i;i=1,2,3)\mapsto (f'_i,g'_i;i=1,2,3)$.\\
The renormalization transformation we define can be also constructed
for the following family of Hamiltonians~:
\begin{equation}
  \label{eq:Hndg}
  H(\u{A},\u{\phi})=\u{\omega}_0\cdot \u{A}+ \frac{1}{2}\u{A}\cdot M\u{A}
  +\u{g}(\u{\phi})\cdot \u{A} +f(\u{\phi}),
\end{equation}
where $M$ is a $3\times 3$ symmetric matrix with non-zero mean-value,
and $\u{g}$ is a three dimensional vector. Applying the
renormalization changes the matrix $M$ into~:
\begin{equation}
\label{eq:rgnd}
{\mathcal{R}}(M)=\frac{\tilde{N} M N}{\mbox{tr }(\tilde{N} M N)},
\end{equation}
where $\mbox{tr }(\tilde{N}MN)$ is the trace of the matrix
$\tilde{N}MN$.  By iterating the map~(\ref{eq:rgnd}), the matrix
converges to $\u{\Omega}^{(3)} \otimes \u{\Omega}^{(3)}$. Moreover,
$\u{g}$ is renormalized into $\tilde{N}\u{g}$ which tends to be aligned
to $\u{\Omega}^{(3)}$ by iteration. Then the
Hamiltonians~(\ref{eq:Hndg}) tend to the degenerate
Hamiltonians~(\ref{eq:Hp}) under the iterations of renormalization. We
expect that the Hamiltonians~(\ref{eq:Hndg}) belong to the same
universality class as the Hamiltonians~(\ref{eq:Hp}).

\section{Renormalization flow}

The renormalization transformation $\mathcal{R}$ has the following
properties~: $\mathcal{R}$ has an attractive integrable fixed point
$H_0$ given by
$$
H_0(\u{A})=\u{\omega}_0\cdot\u{A}
+\frac{1}{2}(\u{\Omega}^{(3)}\cdot\u{A})^2,
$$
and another non-integrable fixed point $H_*$ which lies on the
boundary of the domain of attraction of $H_0$. Outside the closure of
the domain of attraction of $H_0$, the iterations of renormalization
diverge to infinity. The renormalization dynamics has the same
qualitative features as the renormalization for the golden mean torus
for Hamiltonian systems with two degrees of freedom~\cite{chan98b}.
The properties of critical tori are given by the analysis of the
renormalization around the non-trivial fixed point $H_*$.  In
particular, the existence of a non-trivial fixed point for an exact
renormalization would imply self-similarity of critical invariant
tori. The stable manifold of $H_*$ is of codimension one; the
linearized renormalization around $H_*$ has only one eigenvalue of
modulus larger than one~: $\delta \approx 3.4414 $. The value of the
total rescaling coefficient in the actions is $\lambda \approx
11.2726$, and
the rescaling coefficient of time is $\tau +1\approx 2.2469$. \\
We also find a critical fixed cycle with period 7 on the critical
surface.  This periodic cycle is obtained from the non-trivial fixed
point $H_*$ by changing the sign of the Fourier coefficients $f_i$ and
$g_i$. The existence of this cycle is explained by symmetry reasons as
it was done for the critical cycle with period 3 for the golden mean
case~\cite{chan98a}. In particular, it involves the same critical
exponents and scaling factors as the non-trivial fixed point, and 
belongs to the same universality class.

\section{Conclusion}

We have defined two elementary operators ${\mathcal{R}}_P$ and
${\mathcal{R}}_L$. With these operators, we have defined an approximate
renormalization in order to study invariant tori with frequency vector
$\u{\omega}_0=(\tau^2+\tau,\tau,1)$.  The renormalization has a
hyperbolic fixed point with codimension one stable manifold.
Consequently, we expect critical invariant tori with this frequency
vector to be self-similar at criticality. In order to give a firm
basis of this statement, it will be interesting to build an exact
renormalization
transformation without the drastic approximations we used. \\
We notice also that the approximate renormalization we define in this
paper can be generalized to an arbitrary incommensurate frequency
vector $\u{\omega}_0$ given by its Farey sequence, with the operators
${\mathcal{R}}_P$ and ${\mathcal{R}}_L$. For a periodic Farey sequence the
hyperbolic invariant sets are expected to be fixed points, periodic
orbits or strange non-chaotic attractors (such as the spiral mean
torus~\cite{chan99c}). For a non-periodic Farey sequence, this fixed
sets can be strange chaotic attractors, but this point has not been
investigated yet.

\section*{acknowledgments}

We acknowledge useful discussions with G.\ Benfatto, G.\ Gallavotti,
H.R.\ Jauslin, H.\ Koch, and J.\ Laskar.  Support from EC Contract
No.\ ERBCHRXCT94-0460 for the project ``Stability and Universality in
Classical Mechanics'' is acknowledged. CC thanks support from the
Fondation Carnot and from the British Council -- Minist\`ere des
Affaires Etrang\`eres Alliance Program.

\end{document}